\documentclass[
  reprint,       
  aps,            
  prb,            
  superscriptaddress,
  floatfix,
  nofootinbib
]{revtex4-2}

\usepackage{amsmath,amssymb}
\usepackage{graphicx}
\usepackage{bm}
\usepackage{physics}
\usepackage{hyperref}
\usepackage{amsmath}
\usepackage{amssymb}
\usepackage{braket}
\usepackage{amsfonts}
\usepackage{dsfont}
\usepackage{soul}
\usepackage{xcolor}
\usepackage{appendix}
\usepackage{bbold}
\usepackage[compact]{titlesec}
\titlespacing{\section}{0pt}{0.8ex}{0.8ex}
\titlespacing{\subsection}{0pt}{0.4ex}{0.4ex}
\expandafter\def\expandafter\normalsize\expandafter{%
    \normalsize
    \setlength\abovedisplayskip{1.5pt}
    \setlength\belowdisplayskip{1.5pt}
    \setlength\abovedisplayshortskip{1.3pt}
    \setlength\belowdisplayshortskip{1.3pt}
}
\newcommand{\tightfigures}{%
  \setlength{\textfloatsep}{6pt plus 1pt minus 2pt}%
  \setlength{\floatsep}{6pt plus 1pt minus 2pt}%
  \setlength{\intextsep}{6pt plus 1pt minus 2pt}%
  \setlength{\abovecaptionskip}{0pt}%
  \setlength{\belowcaptionskip}{0pt}%
}
\newcommand{\beq}{\begin{equation}}
\newcommand{\eeq}{\end{equation}}
\newcommand{\beqa}{\begin{eqnarray}}
\newcommand{\eeqa}{\end{eqnarray}}
\DeclareMathOperator{\diag}{diag}

\newif\ifshowhidden
\showhiddenfalse
\ifshowhidden
  \newcommand{\hide}[1]{#1}
\else
  \newcommand{\hide}[1]{}
\fi

\begin{document}
\tightfigures
\title{Strong Zero Modes in Supersymmetry-Inspired Quantum Circuits}
\author{Alberto Zorzato}
\affiliation{Institute for Theoretical Physics, University of Amsterdam,
Science Park 904, 1098 XH Amsterdam, The Netherlands
\\QuSoft, Science Park 123, 1098 XG Amsterdam, The Netherlands}

\author{Pietro Richelli}
\affiliation{Kavli Institute of Nanoscience, Delft University of Technology, 
Lorentzweg 1, 2628 CJ, Delft, the Netherlands}

\author{Ji\v{r}\'{i} Min\'{a}\v{r}}
\affiliation{Institute for Theoretical Physics, University of Amsterdam,
Science Park 904, 1098 XH Amsterdam, The Netherlands
\\QuSoft, Science Park 123, 1098 XG Amsterdam, The Netherlands}

\author{Kareljan Schoutens}
\affiliation{Institute for Theoretical Physics, University of Amsterdam,
Science Park 904, 1098 XH Amsterdam, The Netherlands
\\QuSoft, Science Park 123, 1098 XG Amsterdam, The Netherlands}
\date{\today}
\begin{abstract}
\noindent
We investigate discrete dynamics in quantum circuits with gates corresponding to the $S$-matrix of a supersymmetric 1+1D quantum field theory.  
We show that for a brick-wall configuration such circuits support both localized and delocalized dynamically conserved operators known as strong zero modes (SZM), the number of which depends on the parameter regime.
We demonstrate that, while some of the SZM remain localized at boundaries, other SZM propagate ballistically, guided by a choice of circuit parameters.
Such propagation can be explained by a strong Dzyaloshinskii–Moriya term appearing in the dynamics. We describe how to exploit propagating SZM for quantum information transport and discuss the robustness to various types of noise.
\end{abstract}
\maketitle
\emph{Introduction -- }
The rapid development of quantum hardware has stimulated a large bulk of recent work addressing different aspects of quantum dynamics.
Noteworthy examples are Hamiltonian simulations and time-discretization techniques \cite{Lloyd_1996_Science, childsTheoryTrotterError2021, Zhao_2023_PRXQuantum}, the associated transitions to quantum chaos \cite{Heyl_2019_SciAdv, Vernier_2023_PRL}, shortcuts to adiabaticity \cite{Sels_2017_PNAS, Claeys_PRL_2019}, and random unitary quantum circuits, where gates are sampled from a probability distribution \cite{nahumQuantumEntanglementGrowth2017,nahumOperatorSpreading2018,sierantEntanglementGrowth2023,zhouEmergentStatisticalMechanics2019,fisherRandomQuantumCircuits2023,claeysErgodicNonergodic2021,bertiniScramblingRandomUnitary2020}.

The known difficulty to simulate quantum dynamics on a classical computer suggests that this is a field where quantum advantage can be realized. Recent claims in this direction include the digital quantum computation of out-of-time ordered correlation functions \cite{Google_2025_Nature} and the Hamiltonian dynamics of a transverse field Ising model using quantum annealing \cite{King_2025_Science} or discretized time evolution \cite{Kim_2023_Nature} (see, however, \cite{Tindall_2024_PRXQuantum, Begusic_2024_SciAdv}).

Valuable insights on quantum dynamics can be gained if a circuit is endowed with additional structure. This is the case for dual time-unitary circuits \cite{FromDualUnitarity2023} or for circuits constructed as discretized time evolutions of exactly solvable systems \cite{Krajnik2020IntegrableMatrixModels,Rosenberg2024Magnetization,Hudomal2024IntegrabilityBreaking,Paletta2025OpenBoundaryDrivenCircuits,miaoFloquetBaxterisation2024,piroliExactDynamicsDualUnitary2020,mi2022resilientedge,Zadnik2024}. Such structure allows the derivation of analytical results for non-equilibrium steady 
states and correlation functions \cite{Vanicat2018IntegrableTrotterization, Hutsalyuk2025ExactSpinCorrelators}, ballistic spin transport in Floquet setting \cite{Ljubotina2019BallisticSpinTransport} or the formulation of generalized hydrodynamics \cite{Hubner2025GHDCircuits}. Here, the typical way of proceeding is to start from a Hamiltonian $H$ and deriving a discrete Trotter decomposition \cite{suzukiFractalDecompositionExponential1990,suzukiGeneralTheoryHigherorder1992,childsTheoryTrotterError2021} for its evolution operator $U(t)=e^{-iHt}$.

\begin{figure}[htbp]
  \centering
  \includegraphics[width=0.47\textwidth]{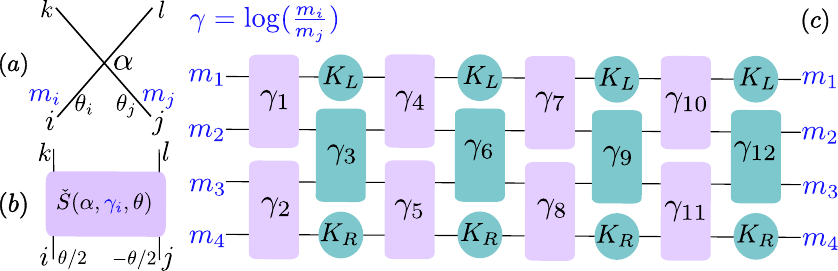} 
  \vskip 1mm
  \caption{\textbf{(a)} Scattering diagram for asymptotic states $i,j,k,l \in\{|b\rangle,|f\rangle\}$, parametrized by interaction strength $\alpha$, rapidities $\theta_i$ and mass ratio $\gamma$. \textbf{(b)} Fundamental 2-qubit gate for $i,j,k,l\in\{|0\rangle,|1\rangle\}$. \textbf{(c)} Brick wall circuit with boundary operators $K_L$, $K_R$, satisfying $[U_F,Q^T]=0$.
  }
  \label{fig:fieldtheorytoqc}
\end{figure}

In this work we follow the inverse path: building on our previous results \cite{richelli2024gfs}, we define a discrete-time evolution operator $U_F$ by choosing the unitary scattering matrix of a (1+1)-dimensional supersymmetric quantum field theory (SUSY QFT) as the fundamental $2$-qubit gate of a quantum circuit of brick-wall type, cf.~Fig.~\ref{fig:fieldtheorytoqc}(c). In the following we refer to $U_F$ as the Floquet operator, although we don't study the effect of driving the system periodically. We find that such circuits feature so-called strong zero modes (SZM), which have their origin in the supersymmetry of the underlying 1+1D QFT. Typically, SZM are localized at system boundaries \cite{FendleyZN_2012,FendleyXYZ_2016,FendleyPara_2016,esslerStrongZeroModes2026,jinTopologicalPrethermalStrong2025,gehrmannExactStrongZero2026,moudgalyaStrongZeroModes2026,vernierStrongZeroModes2024, gehrmann2026exactgeneric, kantha2026randomisingmajorana, olund2025boundary}. In this work we establish the existence of SZMs that propagate ballistically, guided by our choice of circuit parameters.  We exploit this fact to construct protocols for routing quantum information, encoding such information in (pairs of) Majorana modes underpinning the SZM.

\emph{Strong zero modes -- }A working definition of SZMs was provided by Fendley \cite{FendleyXYZ_2016}. In the context of quantum circuits, it assumes an evolution operator $U_F$, and a parity operator $P$. The conditions for $\Psi$ to be a SZM read
\begin{align}
[U_F, \Psi] =0, \quad \{P, \Psi\} = 0, \quad \Psi^2 &\propto {\mathrm 1}. \label{eq:szm}
\end{align}
The first condition states the exact conservation of such operators. We can relax this condition by requiring that the commutator vanishes exponentially in some parameter, leading to \textit{approximate} SZM \cite{moudgalyaStrongZeroModes2026}. The second condition implies spectral pairing between parity-even and parity-odd states, implying that any state $|\psi\rangle$ has a partner $\Psi|\psi\rangle$ evolving with the same quasi-energy. The third property is needed for normalizability.

\emph{Fermionic quantum circuits --} The central object of our protocols is a unitary scattering matrix $\check{S}(\alpha,\gamma,\theta)$, which in the original 1+1D SUSY QFT formulation describes elastic scattering of two particles ($i$ and $j$) with masses $m_{i,j}$ and rapidities $\theta_{i,j}$ interacting with a strength $\alpha$ \cite{richelli2024gfs,schoutensSupersymmetryFactorizable1990}, cf. Fig.~\ref{fig:fieldtheorytoqc}(a). Considering two possible states at the input of the $i$-th channel as $\ket{b(\theta_i)}, \ket{f(\theta_i)}$ (and similarly for $j$), the scattering matrix can be written in the $\{\ket{b(\theta_i)}, \ket{f(\theta_i)}\} \otimes \{\ket{b(\theta_j)}, \ket{f(\theta_j)}\}$ basis, together with its coefficients:
\begingroup
\small
\setlength{\jot}{1pt}
\setlength{\abovedisplayskip}{3pt}
\setlength{\belowdisplayskip}{3pt}
\begin{align}
\lefteqn{\check{S}(\alpha,\gamma,\theta)= }
\nonumber \\
& \frac{1}{2 N} \left[ \sum_{\mu=0}^3 f^\mu X^\mu X^\mu + i (Z I + I Z) + g^+ X Y + g^- Y X \right],
\nonumber\\
&f^{0,3} =
\pm \alpha\,\csch\!\frac{\theta}{2}
+ 2\alpha\,\cosh\!\frac{\gamma}{2}\,\csch\theta,
\label{eq:S} 
\\
&f^1 = f^2 = i,
\qquad
g^\pm =
\sech\!\frac{\theta}{2}
\mp 2\,\csch\theta\,\sinh\!\frac{\gamma}{2},\nonumber\\
&N=\sqrt{2 \alpha^2 (\cosh (\gamma ) \csch^2(\theta )+ \coth (\theta ) \csch(\theta ))+1}. \nonumber
\end{align}
\endgroup
$X^\mu = I, X, Y, Z$, for $\mu=0,\ldots,3$, are the identity and Pauli matrices, respectively, $\theta$ is the difference of the incoming rapidities, $\theta=\theta_i-\theta_j$,
and $\gamma=\log(m_i/m_j)$. Due to the elasticity of the scattering process, both the masses and the rapidities are preserved under $\check{S}$, 
meaning that these labels are simply swapped, $\check{S}: m_{i,j} \rightarrow m_{j,i}$, $\theta_{i,j} \rightarrow \theta_{j,i}$. 
For a given $\alpha$ (which we fix to be the same for all gates in the circuit), the $\check{S}$ gates in the brick-wall circuit are thus uniquely specified by the set $\{m_i\}, \{\theta_i\}$, $i=1,\ldots,L$, of incoming masses and rapidities. 
Throughout, we choose the rapidities as $\theta_i=(-1)^{i-1}\theta/2$.

\begin{figure}[h!]
  \centering
  \includegraphics[width=0.3\textwidth]{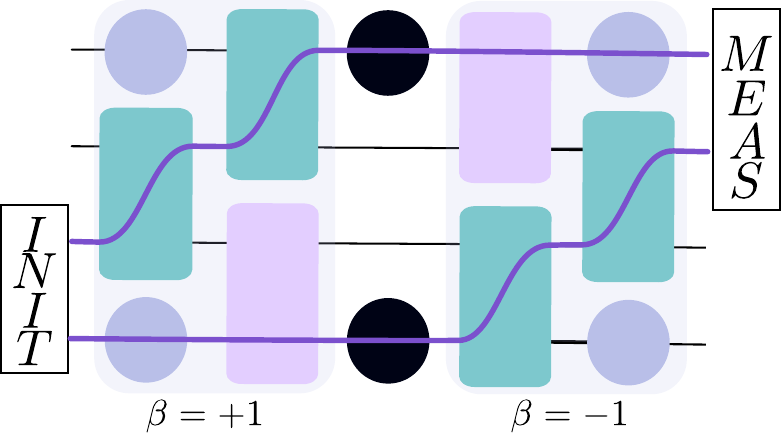}
  \caption{$L=4$ circuit for the transfer one qubit of QI using two spatially separated local SZM. The QI is initialized in qubits 3, 4 and recollected from qubits 1, 2. See SM, \cite{Zorzato2026} for the circuit parameters and the evolution of the SZM.} 
  \label{fig:largegammarouting}
\end{figure}

Working with open boundary conditions (OBC), we need to specify the single-qubit gates $K$ at the edges of the circuit. These flip the incoming rapidity, $\theta_i \rightarrow -\theta_i$, and take the form $K_L=K(\beta,-\theta/2)$, $K_R=K(\beta,\theta/2)$ with
\vspace{1pt}
\beq
K(\beta,\theta)=
\sqrt{2\,\sech(\beta\theta)}\,\diag(k_+,k_-)
\eeq
\vspace{3pt}
\noindent
and $k_{\pm}=\cos[(\pi\pm2i\beta \theta)/4]$. We work with $\beta=1$, but we shall later set $\beta=-1$ in some of the circuit layers to steer the location of the SZM (see Fig.~\ref{fig:largegammarouting}).

The labels $\{m_i\}$, $\theta$ characterizing our brick wall circuits return to their initial values after at most $2L$ layers. We denote the corresponding $2L$-layer unitary by $U_F(\alpha,\{m_i\},\theta)$ (see Fig.~\ref{fig:fieldtheorytoqc}c). 

Viewed as a 2-qubit gate (Fig.~\ref{fig:fieldtheorytoqc}b), the gate $\check{S}$ is a so-called \emph{matchgate}. 
As a consequence, the dynamics associated to our circuits is free fermionic \cite{Valiant2001,Valiant2002,TerhalDiVincenzo2002,JozsaMiyake2008}. 
It is convenient to perform a Jordan-Wigner transformation mapping spins to fermionic annihilation and creation operators $c_j, c^\dag_j$ \cite{JordanWigner1928,LiebSchultzMattis1961,BatistaOrtiz2001} and to define Majorana operators $\chi_j^{A}=c_j^{\dagger}+c_j,\quad \chi_j^{B}=i\,(c_j^{\dagger}-c_j)$ satisfying $\{\chi^\alpha_j,\chi^\beta_k\}=2\delta_{jk}\delta_{\alpha\beta},\quad \alpha,\beta\in\{A,B\}$.

\emph{SZM from global fermionic symmetries --} For PBC, the unitaries $U_F(\alpha,\{m_i\},\theta)$ admit left and right global fermionic symmetries (GFS), which have their origin in the supersymmetry of the underlying QFT,
\begin{equation}  
Q^{L,R}(\{m_i \}, \theta) = N_{L,R} \sum_{i=1}^L \sqrt{m_i} e^{\pm {\theta_i \over 2}} \chi_i^{A,B},
\label{eq:Q}
\end{equation}
where $N_{L,R}$ are normalization constants (see SM, \cite{Zorzato2026} for details). For PBC, both $Q^L$ and $Q^R$ commute with the circuit unitary. 
For OBC, and $\beta=1$, the combination $Q^T=Q^L+Q^R$ survives as GFS \cite{MORICONI1997756,richelli2024gfs,Zorzato2026} and we have
\begin{equation}
[U_F(\alpha,\{m_i\},\theta) , Q^T (\{m_i\},\theta) ]=0.
\end{equation}
The GFS $Q^T$ anti-commutes with parity, $\{Q^T,P\}=0$, 
where $P=\prod_{i=1}^L i \chi_i^A \chi_i^B$, 
and it satisfies $(Q^T)^2 \propto {\mathrm I}$. 
Clearly, $Q^T$ satisfies all properties of Eq.~(\ref{eq:szm}) and it can be identified as a 
\emph{delocalized} SZM, $\Psi_{Q} \equiv Q^T$.

The fact that the left-right symmetric combination $Q^T$ of the GFS charges $Q^L$ and $Q^R$ is a good symmetry for OBC is not surprising. The interesting aspect here is that the underlying free fermionic structure guarantees that there is necessarily a second exact SZM, which we denote as $\Psi_{\rm loc}$, anticipating that in certain parameter regimes it becomes localized. 

\emph{Hamiltonian limit --} 
Throughout the rest of the paper we work in the small $\theta$ regime where
\vspace{2pt}
\begin{equation}
U_F(\alpha,\{m_i\},\theta)
=
\mathbb 1-i\theta H_{\rm eff}(\alpha,\{m_i\})
+\mathcal O(\theta^2).
\label{eq:UF}
\end{equation}
\vspace{2pt}
We consider distinct regimes. 
In the \emph{isotropic} regime, with all masses equal and all $\gamma_i=0$, the Hamiltonian, which can be extracted from the 2-layer Floquet unitary, has been found to represent a critical Kitaev chain \cite{richelli2024gfs,ChepigaMila2023,ChepigaLaflorencie2023}.  
In the \emph{anisotropic} regime we assume a pattern of masses $m$ and $M>m$ on the qubit lines and consider the limit where 
$\gamma = \log(M/m) \rightarrow \infty$. In such a case, the zeroth-order (in $\theta$) of the 2-layer Floquet unitary is not the identity. Such a nontrivial intra-period evolution of operators is commonly referred to as \emph{Floquet micromotion} \cite{GoldmanDalibard2014,Asboth2014,Rudner2013,BukovDAlessioPolkovnikov2015} and it dresses the 2-layer effective Hamiltonian, as explained below.

We define ${\cal M}$ to be the $2L$-dimensional operator space of the linear Majorana operators, spanned by $\bm \chi=(\chi_1^A,\chi_1^B,\cdots,\chi_L^A,\chi_L^B)^\mathsf T$. Since our circuits consist of matchgates, the many-body Floquet unitary $U_F$ induces an orthogonal rotation $O_F\in\mathrm{SO}(2L)$,
\begin{equation}
U_F^\dagger \chi_\mu U_F
=
\sum_{\nu=1}^{2L}(O_F)_{\nu\mu}\chi_\nu .
\label{eq:fermionic_gaussian_rotation}
\end{equation}
In the Hamiltonian limit,
\begin{equation}
O_F(\theta)=\mathbb 1+\theta A+\mathcal O(\theta^2).
\qquad A^\mathsf T=-A .
\label{eq:OF_expansion}
\end{equation}
A general operator $\Psi \in {\cal M}$ is parametrized by a vector of coefficients $\vec v=\{v_1,\ldots,v_{2L}\}$ and can be written as
\begin{equation}
\Psi[\vec v]
=
\sum_{\mu=1}^{2L} v_\mu \chi_\mu
=
\sum_{j=1}^L
\left(
v_{2j-1}\chi_j^A+v_{2j}\chi_j^B
\right).
\label{eq:Psiv}
\end{equation}
With this notation, $U_F^\dagger \Psi[\vec v] U_F=\Psi[O_F\,\vec v]$.
We note that all such linear sums of Majoranas are \textit{odd} under parity, i.e. $\{P,\Psi\}=0$.
This implies that we can identify SZM operators by requiring $O_F\,\vec v=\vec v$, which is equivalent to requiring that $\vec v$ is in the kernel of $A$, $A \vec{v} = 0$, up to first order. 
In the following we track SZM following this procedure.

\emph{Isotropic mass configuration --} For the isotropic mass configuration, $\{m_i\} = \{m,m,\ldots,m\}$, $\gamma=0$ in every 2-qubit gate in the evolution operator.  
In this case we find that $\ker A$ is two-dimensional. For $\alpha=1$ the SZM take the simple form 
\begin{equation}
    \Psi_Q = \sum_{j=1}^L (\chi_j^A + \chi_j^B), \quad
    \Psi_{\rm loc} = {1 \over \sqrt{2}} (\chi^A_L-\chi^B_L).
    \label{eq:Psi_gamma_0}
\end{equation}
Clearly, $\Psi_Q=Q^T$ is uniformly distributed along the chain, while $\Psi_{\rm loc}$ is completely localized on the right edge. We note that setting $\beta=-1$ in the boundary terms $K_L$, $K_R$ will move $\Psi_{\rm loc}$ to the left edge instead. For $\alpha \neq1$, $\Psi_{\rm loc}$ is no longer strictly localized at an edge, but it shows exponential decay into the bulk.

\begin{figure}[h!]
  \centering
  \includegraphics[width=0.44\textwidth]{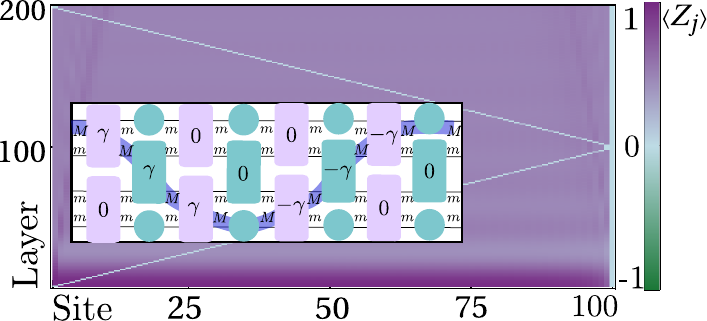}
  \caption{Evolution of local magnetization under $O_F$ for an initial state $\Gamma_0$ and $L=100,\,\alpha=1,\,\theta=0.1,\,\gamma=30$. Here $\{m_i\}=\{M,m,m,\cdots,m,m\}$. SZMs are identified by their magnetization signature: $\langle Z_j\rangle=0$ on the sites where they are localized. The inset shows the circuit diagram for $L=4$.}
  \label{fig:largegammaevo}
\end{figure}

\emph{Anisotropic mass configurations --}
Motivated by the search for SZM, we have considered the following configurations $\{m_i\}$: $\{M,m,M,m,\dots\}$ - staggered masses,
and $\{m, \ldots, m, M, m, \dots,m \}$ - single-heavy-mass, with $M \gg m$ and $\gamma=\log(M/m)$. The single heavy mass $M$ sits on qubit line $k$.  
For such configurations, the $\check{S}$ gates are $\check{S}(\alpha, \gamma,\theta)$ if the incoming masses are $M,m$, and $\check{S}(\alpha, -\gamma,\theta)$ for $m,M$, cf. the examples (for $L=4$) in the insets of Fig.~\ref{fig:largegammaevo}, and Fig.~\ref{fig:largegammarouting}. The explicit construction of the $2L$-layer circuit is in the SM, \cite{Zorzato2026}. 

The two-site gate $\check S(\alpha,\gamma,\theta)$ on qubit lines $(j,j+1)$ acts as an $\mathrm{SO}(4)$ rotation on $\bm \chi^T$. 
In the large $\gamma$, small $\theta$ limit, this rotation simplifies and reduces, up to corrections that are exponentially small in $\gamma$, 
to a permutation of Majorana operators,
\begin{align}
\check S(\alpha,\gamma,0)^\dagger \chi_j^\kappa \check S(\alpha,\gamma,0) &= + \chi_{j+1}^\kappa,\label{eq:chiralpump}\\
\check S(\alpha,\gamma,0)^\dagger \chi_{j+1}^\kappa \check S(\alpha,\gamma,0) &= - \chi_j^\kappa,
\;\;\; \kappa\in\{A,B\}.\notag
\end{align}
\vspace{1.3pt}
The origin of this permutation is the zeroth-order dynamics (in the small $\theta$ expansion) inside one Floquet period. 
To see this, we note that
\begin{align}
\check S_{j,j+1}^{(0)}(\gamma)
& = \check{S}(\alpha, \gamma, 0) \nonumber\\
& = \exp\left[ -i\Phi(\gamma) \left( X_jY_{j+1}-Y_jX_{j+1} \right) \right],
\end{align}
with $\Phi(\gamma)=\frac{1}{2}\arctan\left[\sinh\left(\frac{\gamma}{2}\right)\right]$.

Let us denote by $U_\ell^{(0)}$ the zeroth-order in $\theta$ limit of the unitary $U_{\ell}$ for the $\ell$-th circuit layer. 
Although the full zero-th order is trivial in the sense that $U_{2L}^{(0)}\cdots U_1^{(0)}=\mathbb 1$, the partial products $\mathcal{P}_n^{(0)}=U_n^{(0)}\cdots U_1^{(0)},
$ with $n=1,\ldots,2L-1$, are generally nontrivial. The full Floquet micromotion is defined by the finite-$\theta$ partial products within one period. The operators \(\mathcal P_n^{(0)}\) are their leading-order, small $\theta$ limit, and show that the intra-period motion remains nontrivial even when the zeroth-order stroboscopic evolution closes to the identity. The partial products are periodic over the $2L$-layer cycle and describe how an operator moves within one period before returning to the initial location. In particular, a Majorana operator $\Psi$ evolves intra-period according to $\Psi_n=(\mathcal{P}_n^{(0)})^\dagger \Psi  \,\mathcal{P}_n^{(0)}$. In the large $\gamma$ limit, this micromotion reduces to Eq.~(\ref{eq:chiralpump}). The same micromotion determines $H_{\rm eff}$ as follows. 
Writing  $
U_\ell(\theta)=U_\ell^{(0)}
\left(
\mathbb 1-i\theta h_\ell+\mathcal O(\theta^2)
\right),
$
it follows that
\begin{equation}
H_{\rm eff}
=
\sum_{\ell=1}^{2L}
\left(\mathcal{P}_{\ell-1}^{(0)}\right)^{-1} h_\ell \mathcal{P}_{\ell-1}^{(0)} ,
\end{equation}
as derived in the SM \cite{Zorzato2026}. 

\emph{Approximate SZM --}
We just saw that the generator whose kernel defines the Hamiltonian limit SZM is not the naive sum of the local first-order generators $h_\ell$, but rather of $h_\ell$ dressed by the micromotion. As a consequence, a Majorana operator can be transported through the Floquet period by $\mathcal{P}_n^{(0)}$ before returning to itself after $2L$ layers. Remarkably, the micromotion dressing is responsible for the appearance of additional SZM: in the large $\gamma$ limit the kernel of $A$ becomes four dimensional. In the following we identify these modes for the staggered and single-heavy-mass configurations and denote them with a tilde as $\widetilde{\Psi}_Q$, $\widetilde{\Psi}_{\rm loc}$.

\noindent
\emph{$\{M,m,M,m,\dots\}$}.
In the staggered mass configuration we first identify $\Psi_Q$ and $\Psi_{\rm loc}$. For large $\gamma$ they read 
\begingroup
\small
\setlength{\jot}{1pt}
\setlength{\abovedisplayskip}{3pt}
\setlength{\belowdisplayskip}{3pt}
\begin{align}
\Psi_{\rm loc} &=\frac{1}{2}\left( \chi_{L-1}^A - \chi_{L-1}^B + \chi_L^A - \chi_L^B \right),
\label{eq:exact1}\\
\Psi_Q &= {1 \over \sqrt{L}} \sum_{j\ {\rm odd}} \left( \chi_j^A + \chi_j^B \right).
\label{eq:qtot1}
\end{align}
\endgroup
The remaining two SZM correspond to approximately conserved charges, which become exact for $\gamma\rightarrow\infty$,
\begingroup
\small
\setlength{\jot}{1pt}
\setlength{\abovedisplayskip}{3pt}
\setlength{\belowdisplayskip}{3pt}
\begin{align}
\widetilde\Psi_{\rm loc} & =\frac{1}{2}\left( \chi_{L-1}^A - \chi_{L-1}^B - \chi_L^A + \chi_L^B\right), 
\\
\widetilde{\Psi}_Q & = {1 \over \sqrt{L}} \sum_{j\ {\rm even}} (-1)^{j/2}\left( \chi_j^A + \chi_j^B \right).
\label{eq:qtotapprox1}
\end{align}
\endgroup

\noindent
\emph{$\{m,\ldots, m,M,m, \dots,m\}$}.
In the single-heavy-mass configuration we find
\begingroup
\small
\setlength{\jot}{1pt}
\setlength{\abovedisplayskip}{3pt}
\setlength{\belowdisplayskip}{3pt}
\begin{align}
\Psi_{\rm loc}&= p_L \; \left( \chi_k^A-\chi_k^B \right) + q_L \left( \chi_L^A-\chi_L^B \right), 
\label{eq:exact2}
\\
\Psi_Q &= Q^{T}(\{ m_i\},\theta)  = \frac{1}{\sqrt2}\left( \chi_{k}^A + \chi_{k}^B \right),
\label{eq:qtot2}
\end{align}
\endgroup
with $p_L =1 /\sqrt{2(L^2-2L+2)}$ and $q_L=(L-1) p_L$. For $\gamma$ large, the approximate SZM are 
\begingroup
\small
\setlength{\jot}{1pt}
\setlength{\abovedisplayskip}{3pt}
\setlength{\belowdisplayskip}{3pt}
\begin{align}
&\widetilde\Psi_{\rm loc} = q_L \; \left( \chi_k^A-\chi_k^B \right) - p_L \left( \chi_L^A-\chi_L^B \right),
\label{eq:approxloc2}
\\
&\widetilde\Psi_Q =
{1 \over \sqrt{2L-2}} \bigg[ \sum_{j<k}  \left( \chi_j^A + \chi_j^B \right)-\sum_{j> k}  \left( \chi_j^A + \chi_j^B \right)\bigg].
\label{eq:qtotapproxhole} 
\end{align}
\endgroup

\emph{SZM in Floquet dynamics -- }We now turn to numerical results. Our goal is to visualize the dynamical signature $\langle Z_i\rangle$ of the localized SZM operators identified above. In the Majorana representation, an operator is conserved over one Floquet period when its coefficient vector $\vec{v}$ is a $+1$ eigenvector of $O_F$. We therefore define
\begin{equation}
\Omega:=\ker(O_F-\mathbb 1).
\label{eq:Omega}
\end{equation}
Elements $\vec v$ of $\Omega$ construct the candidate SZM operators Eq.~(\ref{eq:Psiv}).
We emphasize, that in this section, while still using $\theta \ll 1$ corresponding to the Hamiltonian limit, the numerical results are obtained without approximations, which motivates the use of $O_F$ in Eq.~(\ref{eq:Omega}) rather than $A$ in Eq.~(\ref{eq:OF_expansion}). In the numerical calculations we exploit the matchgate structure of our circuits: this allows to use fermionic Gaussian techniques \cite{surace2022fgs} - in every layer of the circuit, the state is fully specified by its covariance matrix with elements $\Gamma_{\mu\nu}=\frac{i}{2}\langle [ \chi_\mu ,\chi_\nu]\rangle$. Denoting the initial state as $\Gamma_0$, the state after $\ell$ circuit layers can be obtained as $\Gamma_\ell=O_\ell\Gamma_{\ell-1} O_\ell^{\mathsf T}$, where $O_\ell$ denotes the Majorana rotation associated with the $\ell$-th layer, with $O_F=O_{2L}\ldots O_1$, cf. Eq.~(\ref{eq:fermionic_gaussian_rotation}) and SM \cite{Zorzato2026}.

To probe the zero-mode sector in state dynamics, we choose two orthonormal vectors $\vec v_1,\vec v_2\in\Omega$. They define two real Majorana zero-mode operators $\Psi_1=\Psi[\vec v_1]$ and $\Psi_2=\Psi[\vec v_2]$.  
We pair $\Psi_1$ and $\Psi_2$ into a Dirac fermion
\begin{equation}
    \eta={1 \over 2}(\Psi_1+ i \Psi_2).
    \label{eq:complexdiracfermion}
\end{equation} 
The associated covariance block is
\begingroup
\small
\setlength{\jot}{1pt}
\setlength{\abovedisplayskip}{3pt}
\setlength{\belowdisplayskip}{3pt}
\begin{equation}
    \Gamma^{(\eta)}=
    \begin{pmatrix}
        0 & 2n_\eta-1 \\
        -(2n_\eta-1) & 0
    \end{pmatrix},
    \qquad n_\eta=\eta^\dagger \eta .
\end{equation}
\endgroup
In the simulations shown here we choose the occupied state $n_\eta=1$, so this block is $J(+1)$, with
\begingroup
\small
\setlength{\jot}{1pt}
\setlength{\abovedisplayskip}{3pt}
\setlength{\belowdisplayskip}{3pt}
\begin{equation}
    J(s)=
    \begin{pmatrix}
        0 & s \\
        -s & 0
    \end{pmatrix}.
\end{equation}
\endgroup
To construct the full covariance matrix, we complete $\vec v_1$, $\vec v_2$ to an orthonormal basis of the Majorana coefficient space, and we write $\mathcal O$ 
for the corresponding orthogonal matrix. We choose the covariance matrix as
\begin{equation}
    \Gamma=J(+1)\oplus J(-1)\oplus J(-1)\oplus\cdots .
\end{equation}
While the first block fixes the occupation of the Dirac fermion $\eta$, the remaining blocks define a reference Gaussian background for the orthogonal directions. Rotating back to the original basis gives
\begin{equation}
    \Gamma_0=\mathcal O\Gamma\mathcal O^{\mathsf T}.
\end{equation}
The state is then evolved by the action of the orthogonal transformations $O_\ell$. 

We follow this procedure for a circuit with a single heavy mass $M$ on the $k=1$ qubit line. For large $\gamma$, the circuit dynamics has two localized SZM, Eqs.~(\ref{eq:exact2}) and (\ref{eq:approxloc2}). We initialize the covariance matrix on the occupied state of the Dirac fermion for the pair $\Psi_{\rm loc}$ and $\widetilde{\Psi}_{\rm loc}$, as described in the above, evolve $\ell$ circuit layers and then plot the local magnetization $\langle Z_j\rangle$, which is the $(2j-1,2j)$ element of the covariance matrix $\Gamma_{\ell}$, see Fig.~\ref{fig:largegammaevo}.

For the occupied ($s=1$) state of the Dirac Fermion constructed out of $\chi_i^A-\chi_i^B$ and $\chi_j^A-\chi_j^B$, both magnetizations $\langle Z_i\rangle$ and $\langle Z_j\rangle$ will be zero. The white traces in Fig.~\ref{fig:largegammaevo} thus track the location of the propagating SZM, and it is clear that, while one sticks to the right boundary, the other propagates ballistically, guided by the location of the heavy mass $M$ in the circuit parameters. In SM, \cite{Zorzato2026} we show the numerical resilience of this magnetization signature under parity-preserving perturbations.

For the staggered mass configuration we find, again in the large $\gamma$ limit, localization of SZM on the right boundary and on the rightmost location where unequal masses $m$ and $M$ come together.

\emph{QI routing protocols -- }
Our findings for ballistic propagation of SZM suggest a variety of protocols for quantum information (QI) routing. A concrete example is the following.
We start with a circuit with a single heavy mass $M$ on qubit line $L-1$. Combining the SZM $\Psi_{\rm loc}$ and
$\widetilde{\Psi}_{\rm loc}$ we can define a Dirac fermion
$
\eta = {1 \over \sqrt{8}}(\chi^A_{L-1}-\chi^B_{L-1} + i (\chi^A_L-\chi^B_L)).
$
For $\alpha=1$, $\gamma$ large and $\theta$ small this fermion commutes with the $2L$-layer Floquet unitary $U_F$ , while under micromotion it evolves 
due to the guided propagation of the SZM. Tuning the circuit as indicated in Fig.~\ref{fig:largegammarouting} (see SM, \cite{Zorzato2026} for details), 
giving circuit unitary $U_{\rm transfer}$, we can arrange that
\begin{equation}
    U_{\rm transfer} \, \eta \sim \zeta \, U_{\rm transfer}
\end{equation}
where
$
\zeta = {1 \over \sqrt{8}}(\chi^A_1+\chi^B_1 + i(\chi^A_2+\chi^B_2)).
$
For an $L$-qubit register, this circuit needs $2L-3$ layers.

In the initial circuit layer, we can encode a qubit state on the basis $\{\ket{1_\eta},\ket{0_\eta}\}$, 
where $\ket{1_\eta}$ is a reference state with $\eta^\dagger \ket{1_\eta} = 0$ and $\ket{0_\eta}=\eta\ket{1_\eta}$. 
The circuit unitary $U_{\rm transfer}$ transforms this state to a corresponding state where $\zeta$ has taken the role of $\eta$. 
This means that one qubit worth of quantum information has been transferred across the circuit, 
from qubit lines $(L-1,L)$ to qubit lines $(1,2)$, where it can be extracted by local measurements of $\zeta$, $\zeta^\dagger$ and the bilinear $\zeta^\dagger \zeta$.

One appeal of schemes like this is that splitting the QI in spatially separate SZM leads to a degree of robustness against noise. This is detailed in SM, \cite{Zorzato2026}, where we analyze the infidelity of a qubit transfer protocol carried by spatially separated SZM due to a 1-qubit noise operator $\exp(i \epsilon Z_j)$. We find that the infidelity scales quadratically with $\epsilon$, in contrast to the linear scaling found when a qubit state is encoded in a single qubit line.

\emph{Outlook -- }In this work we proposed a new mechanism of SZM localization that leverages global fermionic symmetries and Floquet micromotion to transport a conserved charge within one period of evolution of our unitary. Future work should aim at generalizing this localization mechanism in order to understand the minimal algebraic conditions that allow for this behavior.

\emph{Acknowledgements -- }This work was supported by the Dutch Ministry of Economic Affairs and Climate Policy (EZK), as part of the QDNL programme. 

\bibliographystyle{apsrev4-2}
\bibliography{references_gfs}

\clearpage
\onecolumngrid
\appendix

\centerline{\LARGE Supplementary Materials}

\vskip 2mm

\centerline{to the results in the manuscript}

\vskip 2mm

\centerline{\bf  Strong Zero Modes in Supersymmetry-Inspired Quantum Circuits}

\vskip 2mm

\centerline{A.~Zorzato, P.~Richelli, J. Min\'{a}\v{r} and K.~Schoutens}

\vskip 6mm
\begin{center}
\begin{tabular}{ll}
\multicolumn{2}{c}{\bf Table of contents} \\[3.5mm]
\textbf{A} -& Fermionic Gaussian Formalism \\[1mm]
\textbf{B} -& Jordan-Wigner mapping of the GFS charges \\[1mm]
\textbf{C} -& Explicit expressions for the brick-wall circuits \\[1mm]
\textbf{D} -& Explicit representations of brick-wall gates \\[1mm]\textbf{E} -& Micromotion and the Hamiltonian limit \\[1mm]
\textbf{F} -& Effect of noise on the SZM \\[1mm]
\textbf{G} -& Qubit transfer protocol \\[1mm]
\textbf{H} -& Effect of noise on QI carried by SZM
\end{tabular}
\end{center}

\vskip 6mm
\section{Fermionic Gaussian Formalism}
\label{app:gaussian_formalism}
If the evolution operator $U$ for a $L$-qubit system is of so-called fermionic Gaussian form, the evolution of some of its correlators can be computed efficiently, that is, avoiding direct computations in the quantum state space, whose dimension grows exponentially with $L$. By fermionic Gaussian we mean exponential in bilinears of a set of $L$ Dirac fermions or, equivalently, $2L$ Majorana fermions $\chi_\mu$. 

The standard example is hamiltonian evolution for a hamiltonian 
\begin{align}
H=\frac{i}{4}\sum_{\mu,\nu}A_{\mu\nu}\chi_\mu \chi_\nu,
\qquad A^\mathsf T=-A,\qquad A\in\mathbb R^{2L\times 2L},
\end{align}
and time evolution operator $U(t)=e^{-iHt}$. 

For a general fermionic Gaussian unitary $U$, the key relation is the propagation of a single Majorana fermion $m_\mu$,
\begin{align}
U^\dagger \chi_\mu U=\sum_\nu R_{\nu\mu} \chi_\nu,
\qquad
R \in \mathrm{SO}(2L).
\end{align}
Using the orthonormality relation
\begin{align}
\operatorname{Tr}(\chi_\mu \chi_\nu) = 2^L \delta_{\mu\nu}
\end{align}
one finds that the rotation $R$ can be expressed as
\begin{align}
R_{\mu\nu}=\frac{1}{2^L}\operatorname{Tr}(\chi_\mu U^\dagger \chi_\nu U).
\end{align}

\subsection{Covariance matrix and efficient simulation of key observables}

To make the connection to correlators, we introduce the covariance matrix $\Gamma_{\mu\nu}$ for a general $L$-qubit state
\begin{equation}
\Gamma_{\mu\nu}=\frac{i}{2}\,\mathrm{Tr}\!\bigl[\rho\,[\chi_\mu,\chi_\nu]\bigr]
=\frac{i}{2}\langle [ \chi_\mu ,\chi_\nu]\rangle =-\,\mathrm{Im}\,\langle \chi_\mu \chi_\nu\rangle, \qquad \Gamma^{\mathsf T}=-\Gamma .
\end{equation}
Evolving the state with the Gaussian unitary $U$ amounts to the following update of the covariance matrix,  
\begin{align}
  \Gamma \rightarrow R \; \Gamma\, R^{\mathsf T}.
\end{align}

For an open chain with $L=2$ sites the Majorana basis is
\begin{align}
\chi_1=X_1, \qquad
\chi_2=Y_1, \qquad
\chi_3=Z_1X_2, \qquad
\chi_4=Z_1Y_2 .
\end{align}
One quickly finds
\begin{align}
\Gamma_{12}=\frac{i}{2}\langle [ \chi_1, \chi_2] \rangle= \frac{i}{2}\langle [X_1,Y_1] \rangle= \frac{i}{2}\langle 2i \epsilon_{123}Z_1 \rangle=-\langle Z_1 \rangle,
\qquad
\Gamma_{21}=\langle Z_1 \rangle,
\qquad
\Gamma_{34}=-\langle Z_2 \rangle,
\qquad
\Gamma_{43}=\langle Z_2 \rangle.
\end{align}
This implies that the expectation values $\langle Z_j \rangle$ can be efficiently computed through direct update of the covariance matrix. The other correlators encoded in the covariance matrix are
\begin{align}
\Gamma_{13}=\langle Y_1 X_2 \rangle, \qquad \Gamma_{14}=\langle Y_1 Y_2 \rangle, \qquad \Gamma_{23}=\langle X_1 X_2 \rangle, \qquad 
\Gamma_{24}=-\langle X_1Y_2\rangle.
\end{align}
As an example, for the product state $|10\rangle$, the covariance matrix will be
\begin{equation}
\Gamma_{|10\rangle}=
\begin{pmatrix}
 0 & -\langle Z_1 \rangle & \langle Y_1 X_2 \rangle&\langle Y_1 Y_2 \rangle\\
 \langle Z_1 \rangle & 0 & -\langle X_1 X_2 \rangle&-\langle X_1Y_2\rangle\\
 -\langle Y_1 X_2 \rangle&\langle X_1 X_2 \rangle& 0 & -\langle Z_2 \rangle\\
-\langle Y_1 Y_2 \rangle& \langle X_1Y_2\rangle& \langle Z_2 \rangle & 0
\end{pmatrix}=
\begin{pmatrix}
    0&1&0&0\\
    -1&0&0&0\\
    0&0&0&-1\\
    0&0&1&0
\end{pmatrix}.
\label{eq:covmat}
\end{equation}
These results can be generalized to general $L$. Assuming an analogous Jordan-Wigner transformation between local Pauli operators $X_i$, $Y_i$ and $Z_i$ and Majorana fermions (see SM \cite{Zorzato2026}), one expresses the local magnetizations $\langle Z_i \rangle$ as a single entry in the covariance matrix. Similarly, one may express correlators containing an \emph{even} number $2m$ of Majoranas as the Pfaffian of a $2m\times2m$ sub–matrix of $\Gamma$, thereby computing them efficiently.
    
\subsection{Evolution operator $O_F$ for brick-wall Floquet operators}

Object of study in this paper are the evolution operators $U_F$ for brick-wall circuits with GFS, see SM \cite{Zorzato2026} for detailed expressions. 
The building blocks are 2-qubit unitaries $\check{S}_{i,i+1}(\alpha,\gamma,\theta)$ and 1-qubit boundary unitaries $K_i(\beta,\theta)$. Passing to the Majorana basis, we associate Majorana operators $\chi_{2j-1}=\chi_j^A$ and $\chi_{2j}=\chi_j^B$ to the Pauli operators at site $i$, using a Jordan-Wigner transformation (see SM, \cite{Zorzato2026} for details). The unitaries $\check{S}_{i,i+1}(\alpha,\gamma,\theta)$ and $K_i(\beta,\theta)$, give rise to rotations 
$\check{S}^M(\alpha,\gamma,\theta) \in \mathrm{SO}(4)$, acting on $\{\chi_i^A, \chi_i^B, \chi_{i+1}^A, \chi_{i+1}^B\}$ and $K^M(\beta,\theta)\in \mathrm{SO}(2)$, acting on $\{\chi_i^A, \chi_i^B\}$. These operators are then embedded in $\mathrm{SO}(2L)$ acting on the full set $\{\chi_1^A, \chi_1^B, \ldots, \chi_L^A, \chi_L^B\}$. Composing all the bricks one finds the complete update operator $O_F\in\mathrm{SO}(2L)$.

\subsection{Preparation of Gaussian states with zero-modes}
\label{sec:preparation_gaussian_states}

This routine, used in our numerics, constructs an initial Gaussian state by choosing the occupation of a Dirac fermion built from two Majorana zero-mode operators. Let $\vec v_1,\vec v_2\in\Omega\subset\mathbb R^{2L}$ be two orthonormal zero-mode vectors in the Majorana mode space. As stated in the main text, these are coefficient vectors defining the linear Majorana operators
\begin{equation}
\Psi_a=\Psi[\vec v_a]=\sum_{\mu=1}^{2L}(v_a)_\mu \chi_\mu,
\qquad a=1,2.
\end{equation}
The corresponding Dirac fermion is then
\begin{align}
\eta=\frac{1}{2}\left(\Psi_1+i\Psi_2\right),
\qquad
\eta^\dagger=\frac{1}{2}\left(\Psi_1-i\Psi_2\right).
\end{align}

Thus the state preparation fixes is the even bilinear
\begin{equation}
    i \Psi_1 \Psi_2 = 2\eta^\dagger \eta - 1.
\end{equation}
The occupation of this fermionic mode is encoded in the $2\times2$ covariance block
\begin{align}
J(s)=
\begin{pmatrix}
0 & s\\
-s & 0
\end{pmatrix},
\qquad
s=\pm1.
\end{align}
With the covariance convention used here, $s=+1$ and $s=-1$ correspond to the occupied and empty states of the Dirac fermion, respectively. If the orthogonal matrix $O$ collects the Majorana eigenvectors as columns, the covariance matrix of the initial state is constructed as
\begin{align}
\Gamma_0 = O\,\Gamma\,O^{\mathsf T} ,
\end{align}
where $\Gamma$ is block diagonal with the matrix $J(s)$ inserted for the chosen Majorana pair and fixed background signs for all other modes. In this way the routine prepares a Gaussian state in which a specific Dirac fermion, formed by a pair of zero-energy Majorana operators, is either occupied or empty.   

In the special case where the Majorana zero mode $v_1$ is local on qubit $i$, meaning it is of the form $v_1=\alpha c_i + \alpha^* c_i^\dagger$, and where $v_2$ does not have support on site $i$, meaning that $[Z_i,v_2]=0$, one finds that $\langle Z_i \rangle=0$ in the Gaussian states with $s=\pm 1$. This can also understood by noticing that SZM operators $(\prod_{k<j}Z_k)X_i$ or $(\prod_{k<j}Z_k)Y_i$ have null component along $Z_j$, therefore  $\langle Z_i\rangle=0$. We use this observable to track conserved SZM in our numerics.

\section{Jordan-Wigner mapping of the GFS charges}
\label{sec:GFSrepr}
The Global Fermionic Symmetry (GFS) charges $Q_L$ and $Q_R$ for an $L$ site circuit depend on the rapidities $\theta_i$ for each of the lines, and on the parameters $\gamma_i$ of the 2-qubit gate (scattering matrix) $\check S(\alpha,\gamma_i,\theta)$ between lines $i$ and $i+1$. Following our earlier paper \cite{richelli2024gfs} we assume $\theta_i=(-1)^{i-1}\theta/2$.

In the spin basis the general expressions are
\begin{align}
Q^{L}(\{m_i \},\theta) = N_L \sum_{i=1}^L  Q^l_i(m_i ,\theta), \qquad
Q^{R}(\{m_i \},\theta) = N_R \sum_{i=1}^L  Q^r_i(m_i ,\theta),
\end{align}
where $N_L$ and $N_R$ are normalization constants and 
\begin{align}
Q^l_i ( m_i ,\theta)
 =\sqrt{m_i} e^{ \theta_i \over 2} \left( \prod_{j<i}Z_j \right) X_i, 
\qquad
Q^r_i ( m_i ,\theta)
 =\sqrt{m_i} e^{- \theta_i \over 2} \left( \prod_{j<i}Z_j \right) Y_i.
\qquad
\end{align}
The parity strings in these expressions are such that the GFS charges are local in a fermionic representation. Using the Jordan-Wigner transformation
\[
c_j=\frac12\Bigl(\prod_{k<j}Z_k\Bigr)
      (X_j+iY_j),
\qquad
c_j^{\dagger}=\frac12\Bigl(\prod_{k<j}Z_k\Bigr)
      (X_j-iY_j),
\]
\[
\Bigl(\prod_{k<j}Z_k\Bigr)X_j=c_j^{\dagger}+c_j,
\qquad
\Bigl(\prod_{k<j}Z_k\Bigr)Y_j=i\,(c_j^{\dagger}-c_j).
\]
the GFS charges are rewritten as 
\begin{align}
Q^L(\{m_i \},\theta) = N_{L} \sum_{i=1}^L \sqrt{m_i} e^{ {\theta_i \over 2}}  (c_i^{\dagger}+c_i), 
 \quad
Q^R(\{ m_i \},\theta)
 =  N_{R} \sum_{i=1}^L \sqrt{m_i} e^{- {\theta_i \over 2}}  i(c_i^{\dagger}-c_i).
\label{eq:supercharges}
\end{align}
Moving from Dirac to Majorana fermions, we get to the final expression of the GFS charges. We introduce the Majorana algebra
\begin{align}
\chi_j^{A}=c_j^{\dagger}+c_j,\qquad
\chi_j^{B}=i\,(c_j^{\dagger}-c_j), \qquad
\{\chi^\alpha_j,\chi^\beta_k\}=2\delta_{jk}\delta_{\alpha\beta},\qquad \alpha,\beta\in\{A,B\}.
\end{align}
Rewriting the supercharges to these operators leads to
\begin{align}
Q^{L}(\{m_i \}, \theta) = N_{L} \sum_{i=1}^L \sqrt{m_i} e^{ {\theta_i \over 2}} \chi_i^{A},
 \qquad
Q^{R}(\{m_i \}, \theta) = N_{R} \sum_{i=1}^L \sqrt{m_i} e^{- {\theta_i \over 2}} \chi_i^{B},
\end{align}

\section{Explicit expressions for the Brick-wall circuits}
\label{app:pbcobccircuits}

\subsection{PBC}
In the PBC case for the $\{M,m,M,m,\dots\}$ configuration, the Floquet operator is given by the standard brick-wall construction
\begin{align}
U^{P}_F(\alpha,\gamma,\theta)
=
U_{\mathrm{odd}}\,U_{\mathrm{even}} \notag
=\\
\bigg(\prod_{i=1}^{L/2}\check{S}_{2i,2i+1}(\alpha,\gamma,\theta)\bigg)
\bigg(\prod_{i=1}^{L/2}\check{S}_{2i-1,2i}(\alpha,\gamma,\theta)\bigg),
\end{align}
where $L+1\equiv 1$. By construction, the evolution preserves the fermionic charges for all parameter values:
\begin{equation}
[U_F^{P}(\alpha,\gamma,\theta),Q^{L/R}(\gamma,\theta)]=0.
\label{eq:commgamma0}
\end{equation}

\subsection{OBC, $\gamma=0$}

\begin{equation}
U^{O}_F(\alpha,0,\theta)=
K_1(-\theta/2)\bigg(\prod_{i=1}^{L/2-1}\check{S}_{2i,2i+1}(\alpha,0,\theta)\bigg)K_L(\theta/2) \bigg(\prod_{i=1}^{L/2}\check{S}_{2i-1,2i}(\alpha,0,\theta)\bigg),
\end{equation}
satisfies the commutation relation $[U^{O}_F(\alpha,0,\theta),Q^T(0,\theta)]=0$, with $Q^T=Q^L+Q^R$. In all OBC realizations the left and right boundary matrices are evaluated at $\mp\theta/2$ respectively.

\subsection{OBC, $\gamma\neq0$}
In this case satisfying the commutation requires $2L$ layers applied in a layout like Fig. \ref{fig:fieldtheorytoqc}(c). An initial string $\{m_i\}$, sets the values of $\{\gamma_i\}=\{\gamma_1,\gamma_2,\dots,\gamma_{T}\}$ in the $T=L(L-1)$ two-qubit gates throughout the $2L$ layers according to

\begin{align}
(M,M)\ {\rm or}\ (m,m) &:\quad \gamma_i = 0, \notag \\
 (M,m) &:\quad \gamma_i = +\gamma, \notag \\
 (m,M) &:\quad \gamma_i = -\gamma.
 \label{eq:ruleschirality}
\end{align}

For even $L$, the number of two-qubit gates in layer $\ell$ is
$$
n_\ell =
\begin{cases}
L/2, & \ell \ \mathrm{odd},\\
L/2-1, & \ell \ \mathrm{even}.
\end{cases}
$$
Then we define the number of two-qubit gates up to layer $\ell$ as $N_\ell=\sum_{r=1}^{\ell}n_r$,
so that $N_{2L}= T$. We define $G_\ell=\{\gamma_{N_{\ell-1}+1},\ldots,\gamma_{N_\ell}\}$ such that $\{\gamma_i\}=\{G_1,\dots ,G_{2L}\}$. Then the full OBC Floquet unitary can be written compactly as
\begin{equation}
U_F(\alpha,\{\gamma_i\},\theta)
=
U_{2L}(\alpha,G_{2L},\theta)
U_{2L-1}(\alpha,G_{2L-1},\theta)
\cdots
U_2(\alpha,G_2,\theta)
U_1(\alpha,G_1,\theta),
\end{equation}
where products are ordered from right to left in time.
Equivalently, for $k=1,\ldots,L$,
\begin{equation}
G_{2k-1}
=
\{\gamma_{(k-1)(L-1)+1},\ldots,
\gamma_{(k-1)(L-1)+L/2}\},
\qquad
G_{2k}
=
\{\gamma_{(k-1)(L-1)+L/2+1},\ldots,
\gamma_{k(L-1)}\}.
\end{equation}

A diagram of the general case for $L=4$ is shown in Fig.~\ref{fig:fieldtheorytoqc}(c), where the ordering of the entries $\gamma_i$ is indicated explicitly. A specific realization generated by the initial string $\{m_i\}=\{M,m,m,m\}$ is shown in the inset of Fig.~\ref{fig:largegammaevo}. In this case $\{\gamma_i\}=
\{+\gamma,0,+\gamma,0,+\gamma,0,0,-\gamma,-\gamma,-\gamma,0,0\}$, so that, using the layer notation defined above,

\begin{align}
&G_1=\{\gamma,0\},\qquad
G_2=\{\gamma\},\qquad
G_3=\{0,\gamma\},\qquad
G_4=\{0\},\\
&G_5=\{0,-\gamma\},\qquad
G_6=\{-\gamma\},\qquad
G_7=\{-\gamma,0\},\qquad
G_8=\{0\}.
\end{align}
Therefore
\begin{align}
U^{L=4}_F(\alpha,\{\gamma_i\},\theta)
={}&
U_{8}(\alpha,G_8,\theta)
U_{7}(\alpha,G_7,\theta)
U_{6}(\alpha,G_6,\theta)
U_{5}(\alpha,G_5,\theta)
\nonumber\\
&\times
U_{4}(\alpha,G_4,\theta)
U_{3}(\alpha,G_3,\theta)
U_{2}(\alpha,G_2,\theta)
U_{1}(\alpha,G_1,\theta).
\end{align}

\section{Explicit representations of brick-wall gates}
\label{sec:explicitreprpauli}
\subsection{Pauli Basis}
\begin{equation}
\check{S}(\alpha,\gamma,\theta)
= N(\alpha,\gamma,\theta)^{-1}\, M(\alpha,\gamma,\theta)
\end{equation}
\begin{equation}\resizebox{0.95\hsize}{!}{$
M(\alpha,\gamma,\theta)=
\left(
\begin{array}{cccc}
 2 \alpha  \cosh \left(\frac{\gamma }{2}\right) \text{csch}(\theta )+i & 0 & 0 & \alpha  \text{sech}\left(\frac{\theta }{2}\right) \\
 0 & \alpha  \text{csch}\left(\frac{\theta }{2}\right) & 2 \alpha  \sinh \left(\frac{\gamma }{2}\right) \text{csch}(\theta )+i & 0 \\
 0 & -2 \alpha  \sinh \left(\frac{\gamma }{2}\right) \text{csch}(\theta )+i & \alpha  \text{csch}\left(\frac{\theta }{2}\right) & 0 \\
 -\alpha  \text{sech}\left(\frac{\theta }{2}\right) & 0 & 0 & 2 \alpha  \cosh \left(\frac{\gamma }{2}\right) \text{csch}(\theta )-i \\
\end{array}
\right) 
$}
\end{equation}
\begin{equation}
    N(\alpha,\gamma,\theta)=\sqrt{2 \alpha ^2 \cosh (\gamma ) \csch^2(\theta )+2 \alpha ^2\coth (\theta ) \csch(\theta )+1}
\end{equation}
\begin{equation}
K(\beta,\theta)=
\left(
\begin{array}{cc}
 \sqrt{2} \cos \left(\frac{\pi }{4}+\frac{i \beta\theta }{2}\right)
   \sqrt{\text{sech}(\beta\theta )} & 0 \\
 0 & \sqrt{2} \cos \left(\frac{\pi }{4}-\frac{i \beta \theta }{2}\right)
   \sqrt{\text{sech}(\beta\theta )} \\
\end{array}
\right)
\end{equation}

Expanding the bulk gate $\check S(\alpha,\gamma,\theta)$ in the Pauli basis
$X^\mu\otimes X^\nu$ ($\mu,\nu=0,1,2,3$) yields
\begin{equation}
\begin{aligned}
\check S(\alpha,\gamma,\theta)
=\frac{1}{2\, N(\alpha,\gamma,\theta)}
\Bigg[
&\Big(
\alpha\,\csch\!\frac{\theta}{2}
+2\alpha\,\cosh\!\frac{\gamma}{2}\,\csch\theta
\Big)\, I\otimes I
+\, i\,\Big(Z\otimes I
+ I\otimes Z
+X\otimes X
+Y\otimes Y\Big)\\[4pt]
&+\Big(
-\,\alpha\,\csch\!\frac{\theta}{2}
+2\alpha\,\cosh\!\frac{\gamma}{2}\,\csch\theta
\Big)\, Z\otimes Z
+\, i\alpha
\Big(
\sech\!\frac{\theta}{2}
-2\,\csch\theta\,\sinh\!\frac{\gamma}{2}
\Big)\, X\otimes Y
\\[4pt]
&+\, i\alpha
\Big(
\sech\!\frac{\theta}{2}
+2\,\csch\theta\,\sinh\!\frac{\gamma}{2}
\Big)\, Y\otimes X
\Bigg],
\end{aligned}
\end{equation}

\subsection{Majorana Basis}

\begin{equation}
\begin{gathered}
D
=
2\alpha^2\bigl(\cosh\gamma+\cosh\theta\bigr)+\sinh^2\theta,
\qquad
\widetilde D
=
2\alpha^2\csch^2\theta\bigl(\cosh\gamma+\cosh\theta\bigr)+1
=
\frac{D}{\sinh^2\theta},
\\[1mm]
A_\pm
=
\frac{4\alpha^2\cosh\!\left(\frac{\gamma\pm\theta}{2}\right)}{D},
\qquad
E_\pm
=
\frac{\alpha\left(\csch\frac{\theta}{2}\pm\sech\frac{\theta}{2}\right)}
{\widetilde D},
\qquad
F_\pm
=
\frac{2\alpha e^{\pm\gamma/2}\csch\theta}{\widetilde D},
\\[1mm]
C_{++}
=
\frac{
2\alpha^2\sinh\gamma
+
\sinh\theta\bigl(2\alpha^2+\sinh\theta\bigr)
}{D},
\qquad
C_{+-}
=
\frac{
2\alpha^2\sinh\gamma
+
\sinh\theta\bigl(\sinh\theta-2\alpha^2\bigr)
}{D},
\\[1mm]
C_{-+}
=
\frac{
-2\alpha^2\sinh\gamma
+
\sinh\theta\bigl(2\alpha^2+\sinh\theta\bigr)
}{D},
\qquad
C_{--}
=
\frac{
-2\alpha^2\sinh\gamma
+
\sinh\theta\bigl(\sinh\theta-2\alpha^2\bigr)
}{D}.
\end{gathered}
\label{eq:SM-defs}
\end{equation}

Then
\begin{equation}
\check S^M(\alpha,\gamma,\theta)
=
\left(
\begin{array}{cccc}
A_- & -E_- & C_{++} & F_- \\[1mm]
E_+ & A_+ & -F_- & C_{+-} \\[1mm]
C_{--} & F_+ & A_- & -E_+ \\[1mm]
-F_+ & C_{-+} & E_- & A_+
\end{array}
\right).
\label{eq:SM-compact}
\end{equation}

\begin{equation}
K^M(\beta,\theta)
=
\left(
\begin{array}{cc}
\sech(\beta\theta) & \tanh(\beta\theta) \\
-\tanh(\beta\theta) & \sech(\beta\theta)
\end{array}
\right).
\label{eq:KM-real}
\end{equation}

\section{Micromotion and the Hamiltonian limit}
\label{app:floquet-Hamiltonian-defect}

Before specializing to our circuit, it is useful to recall the standard structure of Floquet evolution. For a periodically driven system with period $T$, the evolution can be written as
\begin{equation}
U(t)=F(t)e^{-iH_F t},
\qquad
F(t+T)=F(t),
\qquad
F(0)=1.
\end{equation}
Here $H_F$ is the stroboscopic Floquet Hamiltonian, while $F(t)$ is the micromotion, or kick, operator. The important point is that $H_F$ does not, in general, fully characterize the dynamics. The operator $F(t)$ describes the intra-period motion and can carry physical information that is invisible in the stroboscopic Hamiltonian alone. This is particularly relevant in anomalous Floquet phases and Floquet quantum walks, where boundary or defect modes are tied to the full unitary evolution rather than to a static Hamiltonian description. In the brick-wall circuits considered in this work, this distinction is relevant. The full operator is a product of $2L$ layers,
\begin{equation}
U_F(\theta)=U_{2L}(\theta)\cdots U_2(\theta)U_1(\theta),
\end{equation}
where the bulk layers are built from the brick gates $\check S(\alpha,\gamma_n,\theta)$, $\gamma_n\in \{\gamma_i\}$, and the boundaries are closed by the reflection gates $K_L=K(-\theta/2)$ and $K_R=K(+\theta/2)$. The decorated protocols used in the main text are chosen so that, at zeroth order, $U_F(0)=\mathbb 1$. Thus the full-period Hamiltonian limit is indeed an expansion around the identity,
\begin{equation}
U_F(\theta)
=
\mathbb 1-i\theta H_{\rm eff}
+\mathcal O(\theta^2),
\qquad
H_{\rm eff}
=
i\,\partial_\theta U_F(\theta)\big|_{\theta=0}.
\end{equation}
layer This does not mean, however, that the micromotion is trivial. The point is that the identity appears only after multiplying all zeroth-order layers in the full period,
$$
U_{2L}(0)\cdots U_2(0)U_1(0)=\mathbb 1.
$$
The intermediate products
\begin{equation}
\mathcal{P}_n^{(0)}
:=
U_n(0)\cdots U_2(0)U_1(0),
\qquad
n=1,\ldots,2L-1,
\end{equation}
are generally not equal to the identity. They describe the intra-period micromotion. In our case, the zeroth-order bricks route Majorana operators from bond to bond during the period (see Eq.~(\ref{eq:chiralpump})). The decorated sequence is chosen so that this routing closes after $2L$ layers, returning every Majorana operator to its initial position. Thus the net zeroth-order evolution operator is the identity, but the path taken during the period is nontrivial. This dresses the local hamiltonians appearing in the total operator.  To see this we start from the full decorated operator
\begin{equation}
U_F(\theta) = U_{2L}(\theta)\cdots U_2(\theta)U_1(\theta).
\end{equation}
For the decorated protocol, the zeroth-order period closes:
\begin{equation}
U_F(0) = U_{2L}(0)\cdots U_2(0)U_1(0) = \mathbb 1.
\end{equation}
However, the intermediate products
\begin{equation}
\mathcal{P}_n^{(0)} = U_n(0)\cdots U_2(0)U_1(0), \qquad n=1,\ldots,2L-1,
\end{equation}
are generally not equal to the identity. They are the zeroth-order micromotion. Now write the small $\theta$ expansion of each layer in its own co-rotating frame:
\begin{equation}
U_\ell(\theta)=
U_\ell^{(0)}
\left(\mathbb1-i\theta h_\ell+\mathcal O(\theta^2)
\right),\qquad
U_\ell^{(0)}:=U_\ell(0).
\end{equation}
This is the layer version of the local expansion
\begin{equation}
\check S_{j,j+1}(\alpha,\gamma,\theta) = \check S_{j,j+1}(\alpha,\gamma,0)
\left[
\mathbb 1-i\theta h^{(1)}_{j,j+1}(\alpha,\gamma)
+\mathcal O(\theta^2)
\right].
\end{equation}

Expanding the full product to first order gives
\begin{equation}
U_F(\theta)
=
U_{2L}^{(0)}
\left(
\mathbb 1-i\theta h_{2L}
\right)
\cdots
U_2^{(0)}
\left(\mathbb 1-i\theta h_2\right)U_1^{(0)}\left(\mathbb 1-i\theta h_1\right)+\mathcal O(\theta^2).
\end{equation}

The zeroth-order term is
\begin{equation}
U_{2L}^{(0)}\cdots U_2^{(0)}U_1^{(0)}=\mathbb 1.
\end{equation}

Now consider the first-order contribution from layer $k$. We keep $h_k$ from
layer $k$ and set all other layers to zeroth order. This gives
\begin{equation}
-i\theta\,
U_{2L}^{(0)}\cdots U_{k+1}^{(0)} U_k^{(0)} h_k U_{k-1}^{(0)}\cdots U_1^{(0)}.
\end{equation}
Using
\begin{equation}
\mathcal{P}_{k-1}^{(0)} = U_{k-1}^{(0)}\cdots U_1^{(0)},
\end{equation}
and using the fact that the full zeroth-order product is identity,
\begin{equation}
U_{2L}^{(0)}\cdots U_{k+1}^{(0)}U_k^{(0)}\mathcal{P}_{k-1}^{(0)} =
\mathbb 1,
\end{equation}
we get
\begin{equation}
U_{2L}^{(0)}\cdots U_{k+1}^{(0)}U_k^{(0)} =
\left(\mathcal{P}_{k-1}^{(0)}\right)^{-1}.
\end{equation}
Therefore the first-order contribution from layer $k$ is
\begin{equation}
-i\theta\,
\left(\mathcal{P}_{k-1}^{(0)}\right)^{-1} h_k
\mathcal{P}_{k-1}^{(0)}.
\end{equation}

Summing over all layers, we obtain
\begin{equation} U_F(\theta) = \mathbb 1 - i\theta \sum_{k=1}^{2L} \left(\mathcal{P}_{k-1}^{(0)}\right)^{-1} h_k \mathcal{P}_{k-1}^{(0)} + \mathcal O(\theta^2).
\end{equation}
Comparing this with
\begin{equation}
U_F(\theta) = \mathbb 1-i\theta H_{\rm eff} +\mathcal O(\theta^2),
\end{equation}
we find
\begin{equation}
H_{\rm eff} = \sum_{k=1}^{2L} \left(\mathcal{P}_{k-1}^{(0)}\right)^{-1} h_k \mathcal{P}_{k-1}^{(0)}.
\end{equation}

This is the precise sense in which the Hamiltonian limit remembers the
micromotion. Although the full zeroth-order period is the identity, the
intermediate prefix products $P_k^{(0)}$ are not. These prefix products conjugate the local layer Hamiltonians before they are summed into the effective Hamiltonian.  
The local origin of this micromotion is the chiral Dzyaloshinskii-Moriya rotation contained in the zeroth-order brick. Let
\begin{equation}
C_\gamma:=\sech\frac{\gamma}{2}, \qquad
T_\gamma:=\tanh\frac{\gamma}{2}.
\end{equation}
In the computational basis
$\{|00\rangle,|01\rangle,|10\rangle,|11\rangle\}$, one finds
\begin{equation}
\check S_{j,j+1}(\alpha,\gamma,0)
=
\begin{pmatrix}
1&0&0&0\\
0&C_\gamma&T_\gamma&0\\
0&-T_\gamma&C_\gamma&0\\
0&0&0&1
\end{pmatrix}.
\end{equation}
It acts nontrivially only in the odd-parity sector
$\{|01\rangle,|10\rangle\}$. Equivalently,
\begin{equation}
\check S_{j,j+1}(\alpha,\gamma,0)
=
\exp\left[
-i\,\Phi(\gamma)\,
\left(X_jY_{j+1}-Y_jX_{j+1}\right)
\right],
\end{equation}
with
\begin{equation}
\Phi(\gamma)=\frac{1}{2}\arctan\left(\sinh\frac{\gamma}{2}\right).
\end{equation}
Therefore
\begin{equation}
\Phi(+\infty)=+\frac{\pi}{4},
\qquad
\Phi(-\infty)=-\frac{\pi}{4}.
\end{equation}
At large positive and negative $\gamma$,
\begin{equation}
\check S_{j,j+1}(\alpha,+\infty,0)\big|_{\mathrm{odd}}=
\begin{pmatrix}
0&1\\
-1&0
\end{pmatrix},
\qquad
\check S_{j,j+1}(\alpha,-\infty,0)\big|_{\mathrm{odd}}
=
\begin{pmatrix}
0&-1\\
1&0
\end{pmatrix}.
\end{equation}
Thus the sign of $\gamma$ selects the local chirality of the zeroth-order operator. Since the local zeroth-order brick is not the identity, the local Hamiltonian limit is most naturally extracted in a co-moving frame. By this we mean a reference frame which follows the finite zeroth-order motion. Locally, this amounts to factoring the zeroth-order brick $\check S_{j,j+1}(\alpha,\gamma,0)$, and expanding only the residual $\theta$-dependent part around it. We therefore write
\begin{equation}
\check S_{j,j+1}(\alpha,\gamma,\theta)=
\check S_{j,j+1}(\alpha,\gamma,0)
\left[
1-i\theta h^{(1)}_{j,j+1}(\alpha,\gamma)
+\mathcal O(\theta^2)
\right],
\end{equation}
where
\begin{equation}
h^{(1)}_{j,j+1}(\alpha,\gamma)=
i\,\check S_{j,j+1}(\alpha,\gamma,0)^{-1}
\partial_\theta \check S_{j,j+1}(\alpha,\gamma,\theta)
\Big|_{\theta=0}.
\end{equation}

In this co-moving frame the Hamiltonian density is
\begin{equation}
h^{(1)}_{j,j+1}=
J_\gamma (X_jX_{j+1}+Y_jY_{j+1})+
P_\gamma (X_jY_{j+1}+Y_jX_{j+1})+\mu_L(\gamma) Z_j+
\mu_R(\gamma) Z_{j+1},
\end{equation}
with
\begin{equation}
J_\gamma=-\frac{C_\gamma^2}{4\alpha},\qquad P_\gamma=-\frac{C_\gamma}{4},
\end{equation}
and
\begin{equation}
\mu_L(\gamma)=-\frac{e^{\gamma/2}C_\gamma^2}{4\alpha},\qquad \mu_R(\gamma)=
-\frac{e^{-\gamma/2}C_\gamma^2}{4\alpha}.
\end{equation}
Notice that the antisymmetric chiral term
$X_jY_{j+1}-Y_jX_{j+1}$ is absent from $h^{(1)}_{j,j+1}$
because it has already been absorbed into the finite zeroth-order factor
$\check S(\alpha,\gamma,0)$. 
Using the Jordan-Wigner convention
\begin{equation}
c_j=\frac{1}{2}
\left(\prod_{\ell<j}Z_\ell\right)
(X_j+iY_j),
\qquad
c_j^\dagger=\frac{1}{2}
\left(\prod_{\ell<j}Z_\ell\right)
(X_j-iY_j),
\end{equation}
one has
\begin{equation}
X_jY_{j+1}-Y_jX_{j+1}=
-2i(c_j^\dagger c_{j+1}-c_{j+1}^\dagger c_j).
\end{equation}
Therefore the finite part of the brick is a complex nearest-neighbour hopping phase. In the large-$|\gamma|$ limit, the two signs of $\gamma$ correspond to opposite imaginary hopping phases. In Majorana variables
\begin{equation}
\chi_j^A=c_j+c_j^\dagger,\qquad \chi_j^B=i(c_j^\dagger-c_j),
\end{equation}
the chiral hopping term becomes, up to the overall sign fixed by the
Jordan-Wigner convention,
\begin{equation}
X_jY_{j+1}-Y_jX_{j+1}=-i\left(\chi_j^A \chi_{j+1}^A+\chi_j^B \chi_{j+1}^B\right).
\end{equation}

Thus the large-$|\gamma|$ brick transports the two Majorana species independently, with opposite sign for the two signs of $\gamma$:
\begin{align}
\gamma=+\infty:
\qquad
h^{(0)}_{j,j+1} &\propto-i\left(\chi_j^A \chi_{j+1}^A+\chi_j^B \chi_{j+1}^B\right),\notag
\\
\gamma=-\infty: \qquad h^{(0)}_{j,j+1} &\propto i\left( \chi_j^A \chi_{j+1}^A + \chi_j^B \chi_{j+1}^B\right).
\end{align}

The defect is encoded in the micromotion as a mismatch in the zeroth-order routing circuit parameters. 

\section{Effect of noise on the SZM}
\label{app:noiseSZM}
Using the Fermionic Gaussian Formalism we can introduce parity preserving noise in the evolution of the initial state $\Gamma_0$, with $\{m_i\} = {M,m,m,\dots,m,m}$. We look at two different types of noise, phase flip ($Z$ gate) and next nearest neighbor bit flip ($XX$ gate). Both unitaries are matchgate, therefore they can be inserted in the evolution using the formalism introduced in SM \cite{Zorzato2026}. The perturbation in the circuit are inserted after every layer in the brick work at a random position. 

\begin{figure}[h!]
    \centering
    \includegraphics[width=0.28\linewidth]{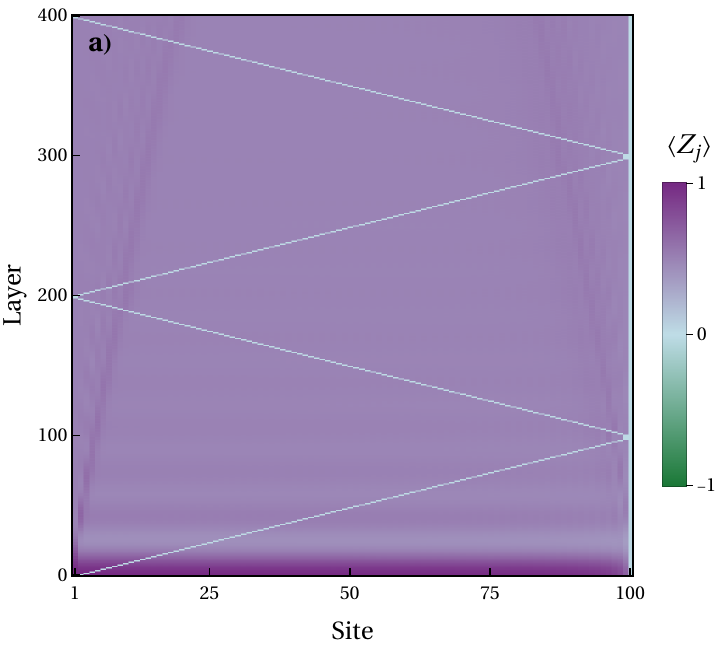}
    \includegraphics[width=0.28\linewidth]{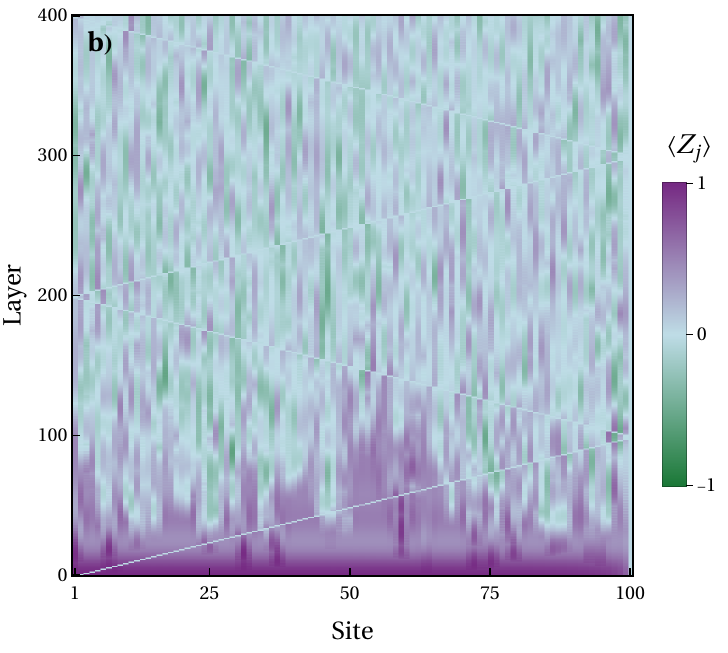}
    \includegraphics[width=0.28\linewidth]{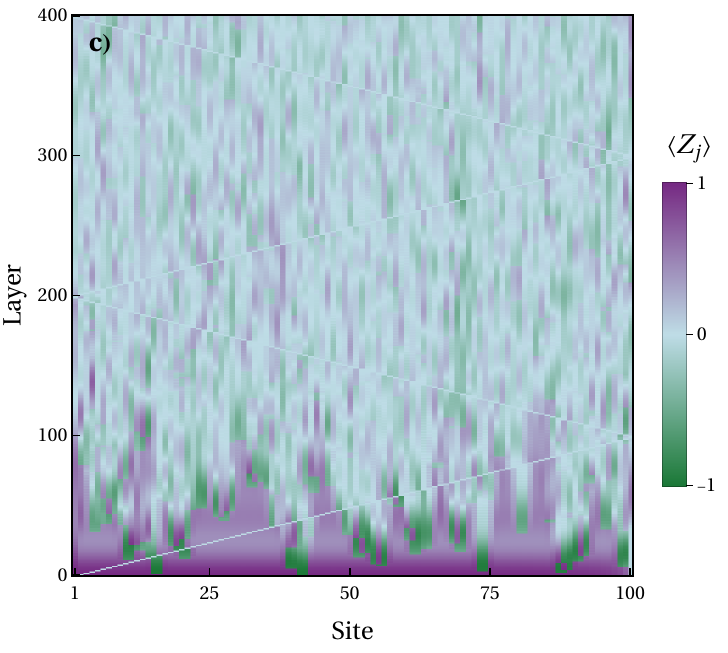}\\
    \includegraphics[width=0.84\linewidth]{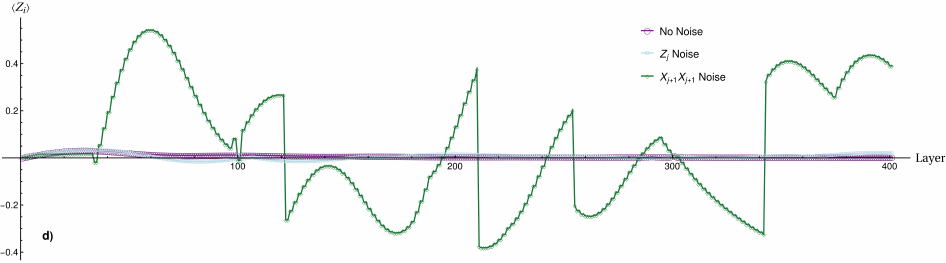}\\
    \includegraphics[width=0.85\linewidth]{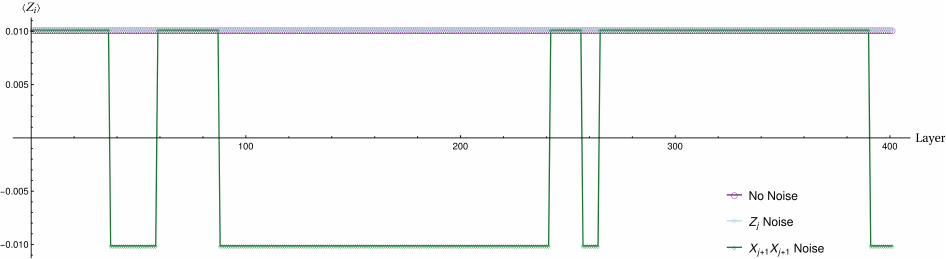}
    \caption{Study of the SZM signature for different parity preserving noises for the evolution of $\Gamma_0$ for $L=100,\,\alpha=1,\,\theta=0.1,\,\gamma=30$, $\{m_i\}=\{M,m,m,\cdots,m,m\}$ and $k=1$. a) Reproduced the graph in fig \ref{fig:largegammaevo}, b) The plot shows the evolution of the local magnetization under phase flip perturbations. c) It shows the evolution under nearest neighbor bit flip perturbations. d) It shows the local magnetization on the right boundary sites. e) It shows the local magnetization for the propagating SZM.}
    \label{fig:single_error_SZM}
\end{figure}

This preliminary analysis, clearly shows the resilience of the protocol under the effect of phase flip perturbation. While it highlights the high susceptibility of the SZM under bit flip errors. Moreover it is important to notice how the scale of the oscillation of the signature $\langle Z_i\rangle$ differ of one order of magnitude. This shows how the propagating SZM could be more resilient to noise than the boundary SZM.

\section{Qubit transfer protocol}
\label{app:trans}

We detail the structure of the circuit (shown in Fig.~\ref{fig:largegammarouting} for $L=4$) that can transfer a qubit of QI across. In words, the idea is that the qubit is encoded in the first circuit layer using the Dirac fermion
$$
\eta = {1 \over \sqrt{8}}(\chi^A_{L-1}-\chi^B_{L-1} + i (\chi^A_L-\chi^B_L)).
$$
The circuit is such that
\begin{equation}
    U_{\rm transfer} \, \eta \sim \zeta \, U_{\rm transfer},
\end{equation}
where
$$
\zeta = {1 \over \sqrt{8}}(\chi^A_1+\chi^B_1 + i(\chi^A_2+\chi^B_2)),
$$
so that the qubit can be extracted by measurements on qubit lines 1 and 2.

The evolution from $\eta$ to $\zeta$ is done in three steps. 
\begin{enumerate}
    \item Starting from mass configuration $\{m,m, \ldots,M,m\}$, $(L-2)$ circuit layers guide the heavy mass from line $L-1$ to line $1$, so that $\eta$ evolves to $\eta'={1 \over \sqrt{8}}(\chi^A_1-\chi^B_1 + i (\chi^A_L-\chi^B_L))$.
    \item Circuit layer $L-1$ has 1-qubit gates $F_1$ and $F_L$ at lines 1 and $L$, which are such that
    $ F_1 F_4 \eta' = \zeta' F_1 F_4 $, with $\zeta'={1 \over \sqrt{8}}(\chi^A_1+\chi^B_1 + i (\chi^A_L+\chi^B_L))$.
    \item The layers $L$ through $2L-3$ are such that one SZM is stuck at line 1, while the other evolves from line $L$ to line $2$. To achieve this, we need to change the sign in the boundary gates, setting $\beta=-1$, and re-initialize the mass configuration in circuit line $L$ as $\{m,m,\ldots,m,M\}$. This fixes the SZM ${1 \over \sqrt{2}} (\chi_1^A+\chi_1^B)$ at line 1, while it propagates the SZM ${1 \over \sqrt{2}} (\chi_L^A+\chi_L^B)$ from line $L$ to line $2$. In all, it propagates $\zeta'$ to $\zeta$.
\end{enumerate}

\section{Effect of noise on QI carried by SZM}
\label{app:noiseQI}

Our main text discusses QI routing protocols, where quantum information is carried by localized SZM that are conserved under the evolution $U_F$. In the setup with single heavy mass, 
$\{m_i\} = \{ m, \ldots, m,M,m, \dots,m \}$, and for parameters $\alpha=1$, $\theta \ll1$, $\gamma \gg 1$, the localized SZM are of the form
\begin{align}
   \Psi_{\rm loc}^k =  {1 \over \sqrt{2}} (\chi_k^A-\chi_k^B) &= {1+i \over \sqrt{2}}(c_k - i c^\dagger_k)
\end{align}
with one of the site indices $k=L$ and the other propagating ballistically under the micromotion in a $2L$ layered circuit. Using the complex fermion 
$\eta = \Psi_{\rm loc}^{k_1} + i \Psi_{\rm loc}^{k_2}$, one can encode a single qubit in the state
\begin{align}
    \ket{\theta,\phi} = \cos({\theta \over 2}) \eta\ket{1_\eta} + \sin({\theta \over 2}) e^{i \phi} \ket{1_\eta}, 
\end{align}
where $\eta^\dagger \ket{1_\eta} = 0$.

Let us now investigate the effect of a 1-qubit noise operator $N^Z_j=\exp(i \epsilon Z_j)$
on, say, qubit $j=k_1$ carrying the SZM $\Psi_{\rm loc}^{k_1}$. It is quickly found that
\begin{align}
    N^Z_k {\chi_k^A - \chi_k^B \over \sqrt{2}} (N^Z_k)^{-1} =  \cos(2\epsilon) {\chi_k^A - \chi_k^B \over \sqrt{2}} - \sin(2\epsilon) {\chi_k^A + \chi_k^B \over \sqrt{2}}.
\end{align}
This implies that the error in quantum memory or qubit transfer protocols due to such noise terms is quadratic in $\epsilon$. This is in contrast to a naive scheme where the quantum information is encoded in a single qubit: in such schemes the effect of 1-qubit noise terms typically scales linearly in $\epsilon$. Extracting Bloch angles $\theta_N$, $\phi_N$ from the state $N^Z_{k_1} \ket{\theta,\phi}$ we find
\begin{align}
\theta_N = \theta + {2 \over \tan(\theta)} \epsilon^2 + \ldots, \qquad
\phi_N &= \phi - \cos(2 \phi) \epsilon^2 + \ldots.
\end{align}

\end{document}